\documentstyle[12pt,epsfig]{article}
\let\section=\subsection  \let\subsection=\subsubsection

\def\be{\begin{equation}}
\def\ee{\end{equation}}
\def\bea{\begin{eqnarray}}
\def\eea{\end{eqnarray}}
\def\Bphi{\mbox{\boldmath $\Phi$}}

\begin{document}
\begin{center}
{\large \bf The Casimir energy of skyrmions\\ in the
2+1-dimensional $O(3)$-model
}\\[5mm]
H. Walliser and G. Holzwarth \\[5mm]
{\small \it 
Fachbereich Physik, Universit\"at Siegen,  
D57068 Siegen, Germany} 
\end{center}

\begin{abstract}\noindent
One-loop quantum corrections to the classical vortices in 
2+1 dimensional $O(3)$-models are evaluated. Skyrme and Zeeman potential
terms are used to stabilize the size of topological solitons.
Contributions from zero modes, bound--states and scattering 
phase--shifts are calculated for vortices with winding index $n=1$ and
$n=2$.
For both cases the S-matrix shows a pronounced series of resonances for
magnon-vortex scattering in analogy to the well-established baryon
resonances in hadron physics, while vortices with $n>2$ are already 
classically unstable against decay. 
The quantum corrections destabilize the 
classically bound $n=2$ configuration. Approximate independence of
the results with respect to changes in the renormalization scale is
demonstrated.  
\end{abstract} 

\bigskip
\leftline{PACS 12.39.Dc, 12.39.Fe, 75.30.Ds, 75.50.Ee}  
\leftline{Keywords: skyrmions, $O(3)$-model}


\section{Introduction}

Effective field theories have found increasing interest 
as powerful tools for describing the dynamics of physical systems
where global symmetries are spontaneously broken and continuous order
parameter fields represent the relevant low-energy degrees of freedom.
Depending on space dimensionality and the manifold on which the fields
live classical localized static solutions may fall into topologically
distinct classes characterized by integer winding numbers, 
carrying energy, momentum and internal properties which suggest their
interpretation as particle-like excitations of the uniform ground
state. Their spatial size is determined by scaling
properties of different competing terms in the effective lagrangian.

Quantization of effective field theories allows to assign proper
quantum numbers to quasi-particle properties, identify
excited states of quasi-particles, and calculate loop
corrections to the classical results for observable quantities,
which may be crucial for experimental verification. The evaluation of
loop corrections necessarily brings about the need for renormalization.
Although, generally effective field theories might be
non--renormalizable in the strict sense, they still may be
renormalized order by order in terms of a gradient expansion.
A well-known example is chiral perturbation theory (ChPT), applied
successfully in hadron physics. In the vacuum sector such an expansion
is truncated by allowing only for external momenta small compared to the
underlying scale of the theory. However in the soliton sector,
the soliton itself constitutes "external" fields with gradients
comparable to the scale of the theory which cannot be made small
by definition. Therefore, in dealing with solitons, the problem
of truncating the expansion
can only be solved by the ad hoc assumption that the renormalized
couplings of the higher gradient terms are small.

A prominent example for the application of this program are 3D-$SU(N_f)$
skyr\-mions in $N_f$-flavor meson fields, where the topological charge 
is identified with baryon number, with impressive success for baryonic
properties, resonances, and meson-baryon dynamics.  
Similarly, the conjecture to consider 2D-$O(3)$ spin textures as charged
quasi-particles in ferromagnetic quantum Hall systems~\cite{SKK93}
and antiferromagnetic high--temperature superconductors~\cite{dpw88},
with topological winding density identified with the deviation of the
electron density from its uniform background value, suggests a 
corresponding investigation. 
For magnons and vortices with unity charge such attempts have
been presented for ferromagnets~\cite{a97} and 
antiferromagnets~\cite{r89}. The field theoretical approach 
forwarded here is rather
related to the antiferromagnet due to its preserved time--reversal
invariance which makes the analysis fairly similar to what applies
to relativistic systems. The ferromagnet where time--reversal
invariance is broken would require the consideration of the
Landau--Lifshitz dynamics.
The evaluation of the Casimir energies involves bound--state
energies and
a sum over scattering phase--shifts~\cite{dhn74}. 
This provides also complete 
information about resonant excited states in the continuum
which in the 3D-$SU(N_f)$ case successfully describe well-established
baryon resonances.

For the general outline of the necessary steps and technique we discuss
the case where
the second-order exchange energy is taken in the time--reversal
invariant form of the non--linear sigma model as it naturally appears in
relativistic field theories. In two dimensions it is scale invariant
and therefore irrelevant for the spatial extent of the static 
localized solution. To fix the soliton size two more terms are needed.
We use the standard fourth-order Skyrme term, and a symmetry-breaking
coupling of the $O(3)$-order parameter to an anisotropy field. 
The static part of the Skyrme term represents a local
approximation to the Coulomb energy of the charged excitations; in our
field-theoretic example we keep also the time-derivative parts of it.
So our discussion will mainly serve a demonstrative purpose as a model
field theory. 
 
We shall specifically address the question of the binding energy of
doubly charged quasi particles. 
For 3D-$SU(N_f)$ skyrmions this question has a long history since it was
discovered that in the winding number $B=2$ sector the static
lowest-energy solution is bound with respect to decay into two separate
$B=1$ skyrmions and displays only axial symmetry in contrast to the
radially symmetric hedgehog skyrmions. There has been much discussion
about the physical relevance (in the deuteron) of such a torus
configuration,
and it was speculated that quantum corrections might reverse the sign
of the binding energy $E(B=2)-2E(B=1)$. Evaluation of loop
corrections to the 3D-$SU(2)$ $B=2$ torus is a formidable task which to our
know\-ledge has not yet been achieved. It is interesting that the same
situation occurs for the 2D-$O(3)$ skyrmions: the classical $n=2$
solution shows a ring-like density distribution and it is bound with
respect to decay into two individual $B=1$ skyrmions. In this case,
however, the evaluation of loop corrections for both configurations is
of comparable complexity, and we will show that they in fact reverse
the sign of the binding energy.

Actually, there is an ongoing discussion~\cite{LLKS97,NK97} 
about the binding of 2-skyrmions also in condensed matter applications. 
However, it should be stressed that for ferromagnetic
systems the time-dependent part of the effective lagrangian has to be
replaced by the T-violating Landau-Lifshitz form with only one time
derivative. 
For antiferromagnets 
the time derivative coupling of the staggered
spin to the magnetic field, which does not contribute to the static
stabilization should be included~\cite{b98,h98,rs99}. 
And, in any case, for quantitative
conclusions, the non--local character of the Coulomb energy should be
respected~\cite{mm97}.

\section{Static solutions}

The Lagrangian of the $O(3)$ model in $2+1$ dimensions
is conveniently written in terms of the $3$-component field
$\Bphi$ satisfying the constraint $\Bphi \cdot \Bphi =1$,
\be\label{lag}
{\cal{L}} = \frac{f^2}{2} \partial_\mu \Bphi \partial^\mu \Bphi
- \frac{1}{4e^2} ( \partial_\mu \Bphi \times \partial_\nu \Bphi )^2 
- f^2 m^2 (1-\Phi_3)    \, . 
\ee
The three terms represent the non--linear sigma 
(N$\ell \sigma$) model, the Skyrme and the
potential term. There exists a conserved topological current
\be\label{top}
T^\mu = \frac{1}{8 \pi} \epsilon^{\mu \nu \rho} \Bphi
( \partial_\nu \Bphi \times \partial_\rho \Bphi ) \, , \qquad
\partial_\mu T^\mu = 0
\, .
\ee
The radially symmetric hedgehog ansatz, written
in polar coordinates $(r,\varphi)$
\be\label{hedgehog}
\Bphi = \left( \begin{array}{c}
\sin F(r) \cos n \varphi   \\ 
\sin F(r) \sin n \varphi     \\
\cos F(r) 
\end{array} \right) \, , 
\ee
corresponds to winding number
\be\label{winding}
\int d^2 r T^0 = \frac{n}{2}[\cos F(\infty) - \cos F(0)] = n
\, , \qquad
T^0 = - \frac{n F^\prime \sin F}{4 \pi r} 
\ee
and the soliton's size may be defined according to
\be\label{radius}
<r^2>_n  \, = \frac{1}{n} \int d^2 r \, r^2 T^0 =
-\frac{1}{4\pi} \int d^2 r \, r F^\prime \sin F
\, . 
\ee
It is convenient to absorb the length $1/\sqrt{fem}$
which sets the scale for the size of localized structures
into the dimensionless spatial coordinate $x=\sqrt{fem} \, r$.
The static energy functional connected with the lagrangian
(\ref{lag}) is then given by
\be\label{static}
E^{class}_n  =  \frac{f^2}{2} \int d^2 x \left[ F^{\prime 2}
+ \frac{n^2 s^2}{x^2} + a \frac{n^2 F^{\prime 2} s^2 }{x^2}
+ 2 a (1-c) \right]  
 =  4\pi f^2 E_n^0 (a) \, , 
\ee
with the abbreviations $s=\sin F$ and $c= \cos F$.
Technically, $a=m/fe$ is the only nontrivial parameter, 
while $4 \pi f^2$ sets the overall energy scale. We shall therefore present 
results for different values of $a$. The limit
$a \to 0$ describes the pure Belavin--Polyakov (BP) solution~\cite{bp75}. 
The opposite limit, $a \to \infty$, is technically also well
defined; it describes a system without spin-spin aligning force.

The Euler--Lagrange equation following from the variation 
of the energy functional
\be\label{stability}
\frac{1}{x} \left(x F^\prime \right)^\prime - \frac{n^2 sc}{x^2}
+ a \frac{n^2 s}{x} \left( \frac{F^\prime s}{x} \right)^\prime 
- a s = 0
\ee
is a second order non--linear differential equation which is solved
subject to the boundary conditions
\be\label{boundary}
F (0) = \pi \, , \qquad  F (x \to \infty) = 0 
\, .
\ee
In principle the boundary condition at the origin 
according to (\ref{winding}) could be
$F(0) = \nu \pi$ with $\nu=1,3,\dots$ an odd multiple of $\pi$,
however it turns out that the solution with $\nu=1$ has the
smallest static energy.
\begin{figure}[htbp]
\begin{center}
\begin{tabular}{ll}
\epsfig{figure=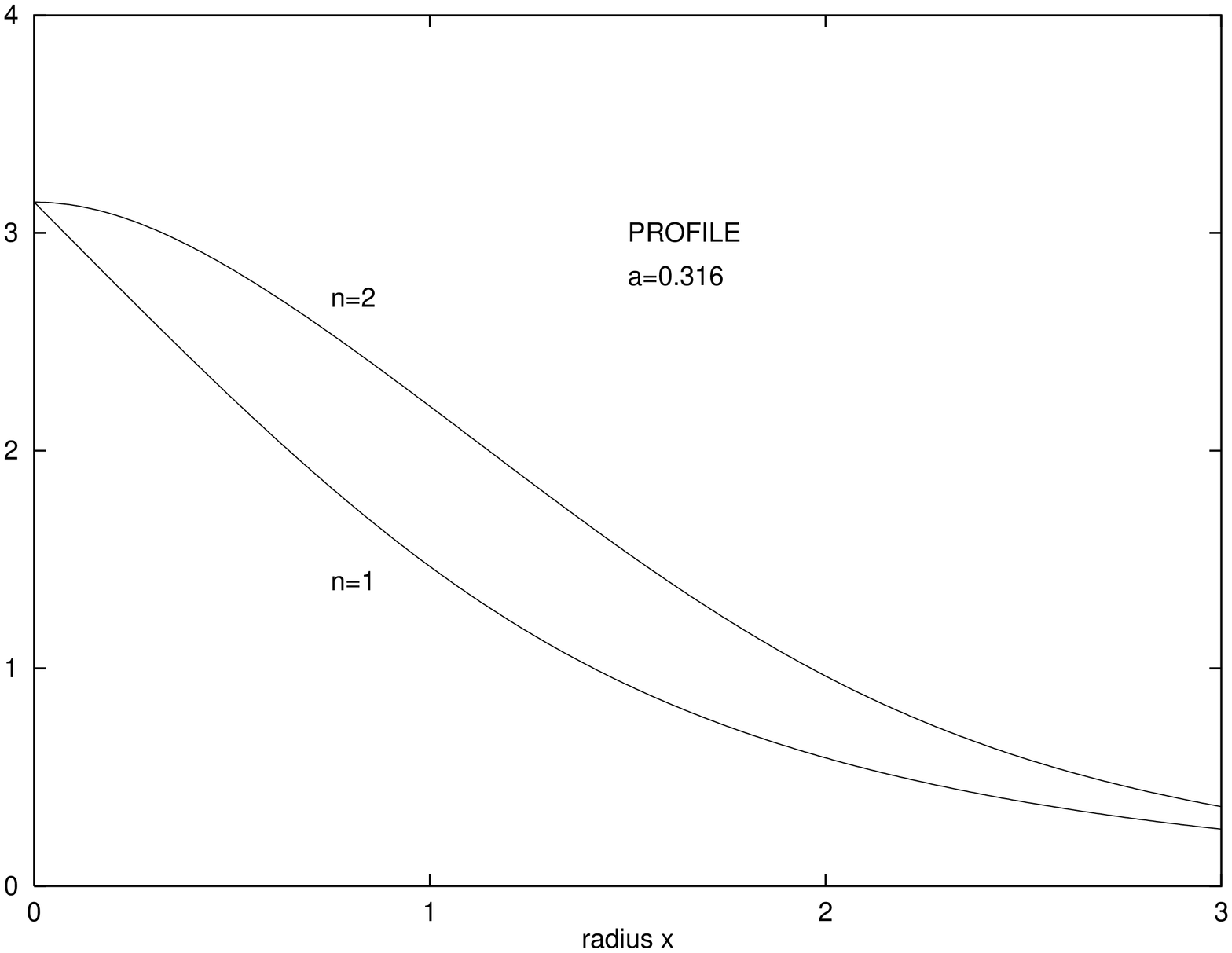,width=5cm,angle=270}
&
\epsfig{figure=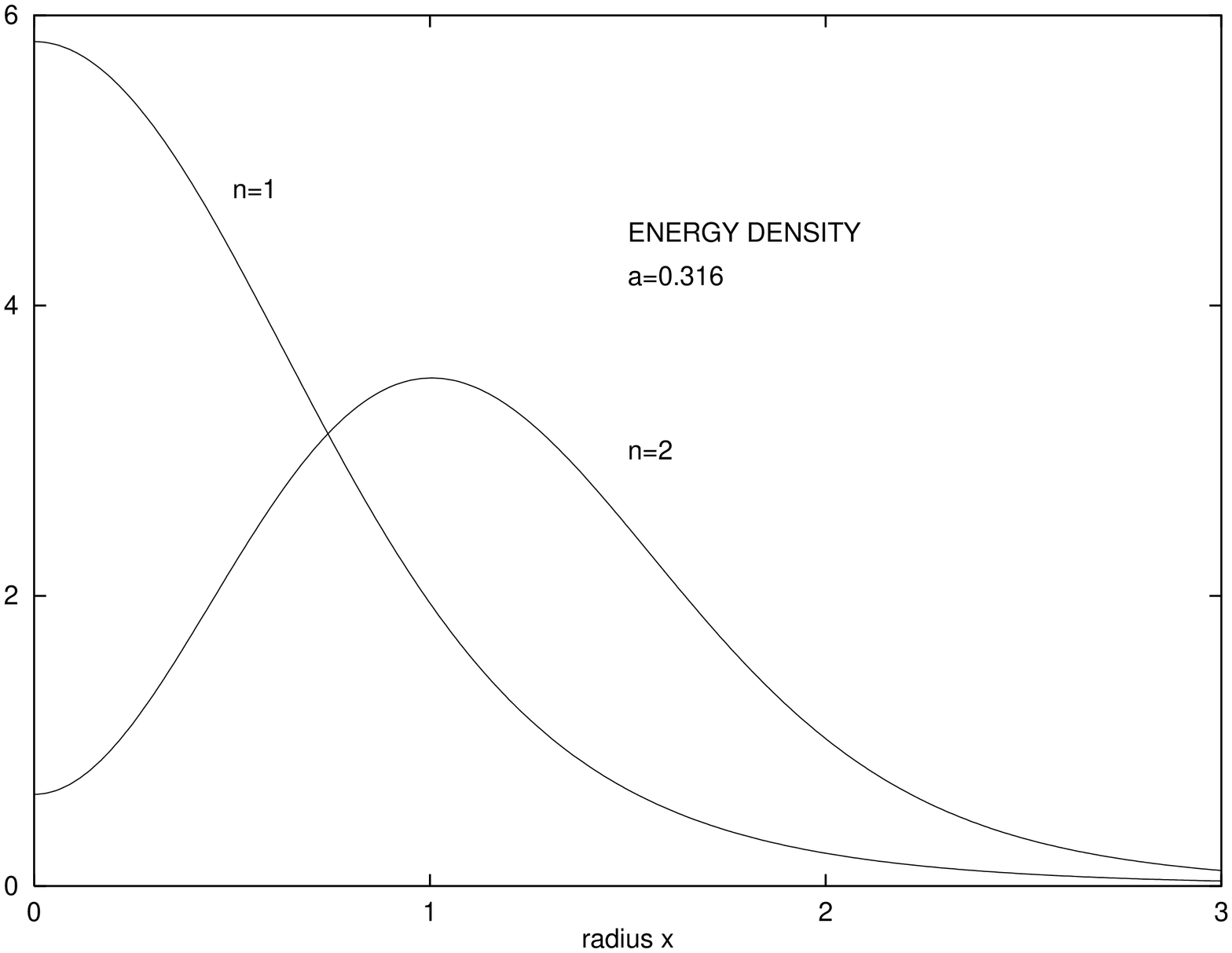,width=5cm,angle=270}
\end{tabular}
\protect\caption{Hedgehog profiles and energy densities for the
skyrmions with $n=1$ and $n=2$. The parameter $a^2=0.1$ was
used.
}
\end{center}
\end{figure}

The hedgehog profiles and the corresponding energy densities for
$a^2=0.1$ are shown in Fig. 1 for the 1- and 2-skyrmions.
It is noticed that the maximum energy density for the $n=2$ soliton
is not located at the
origin but in a spherical shell of radius $x \simeq 1$. 

The classical energies and square radii are given in Table \ref{tab_1}.
\begin{table}
\begin{center}
\parbox{10cm}{\caption{\label{tab_1}
Classical soliton energies and square radii
for the stable 1- and 2-skyrmions
for various parameters $a$.
}} 
\begin{tabular}{|c|cccccc|}
\hline
$a$    & $0.0$  &  $0.01$  &  $0.0316$  & $0.1$ & $0.316$ & $1.0$ \\ 
\hline
$E_1^0$     & $1.00$ & $1.03$  & $1.08$   & $1.21$ & $1.56$ & $2.57$   \\
$E_2^0$     & $2.00$ & $2.03$  & $2.10$   & $2.30$ & $2.94$ & $4.85$   \\
\hline
$<x^2>_1$   & $\infty$ & $2.44$ & $2.11$  & $1.80$   & $1.57$ & $1.42$   \\
$<x^2>_2$   & $ \pi  $ & $3.11$ & $3.06$  & $2.99$   & $2.78$ & $2.76$   \\
\hline
\end{tabular}
\end{center} 
\end{table}
In the two limiting cases, $a \to 0$ and $a \to \infty$, the
differential equation (\ref{stability}) may be solved 
analytically (see appendix). The corresponding static energies 
and radii are
\bea\label{limit}
E_n^0=n \, , \qquad \quad & \, \qquad 
<x^2>_n=\sqrt{\frac{2n}{3} (n^2-1)}
\cdot \frac{\pi/n}{\sin (\pi/n)} 
\qquad \qquad & a<<1 \nonumber \\
E_n^0=4n/3 \cdot a \, , \, & <x^2>_n=4n/3 
\qquad \qquad \qquad \qquad  & a>>1  
\, .
\eea
It is noticed that for the BP solution with $a \to 0$
the 1-skyrmion's radius diverges which is due to the particular
choice of the potential term in (\ref{lag}). This divergence
is perceptible only for extremely small values of $a$, for
$a \stackrel{>}{_\sim} 0.001$ the radius for the 1-soliton is
still smaller than that of the 2-soliton which stays finite
in the limit $a \to 0$ (for a more thorough discussion of this
problem we refer to the appendix).  
The dependence of the energies on the dimensionless parameter
$a$ is shown in Fig. 2. Independently from this parameter 
$E_2^0 \le 2E_1^0$ always holds, the 2-skyrmion is classically
stable against decay into two 1-skyrmions. 
\begin{figure}[htbp]
\centerline{\epsfig{figure=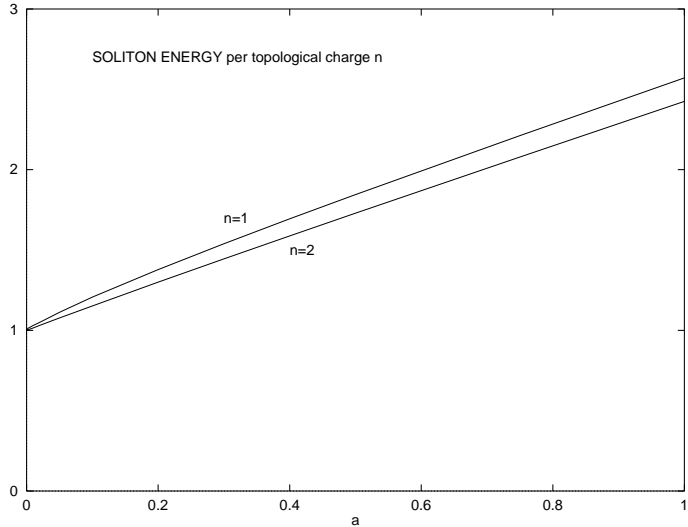,width=7cm,angle=270}}
\protect\caption{
Classical soliton energies for $n=1$ and $n=2$ as functions 
of the dimensionless parameter $a$. For better comparison
the energies plotted are divided by the corresponding topological
charge. The 2-skyrmion is classically stable against decay 
into two 1-skyrmions.
}
\end{figure}
Hedgehog solutions with higher topological
charge, $n \ge 3$, are not stable (see subsection 3.3 
for their stability)
and do not represent the minimum energy configuration of the
corresponding sector. The 3-soliton's minimum energy configuration
is not rotationally symmetric, it is a linear molecule which
consists of three distorted 1-solitons. Higher multi--solitons
are then supposedly molecules of stable 2-skyrmions (and
1-skyrmions in case of $n$ odd)~\cite{psz95}.  This
picture may change fundamentally when the Casimir energies
are taken into account.

\section{Small amplitude fluctuations}

The evaluation of quantum corrections is based on the normal modes
of the classical configurations. With the appropriate boundary
conditions these normal modes describe excited localized (bound) states,
and the scattering of the vacuum fluctuations (mesons or magnons)
off the soliton backgound. (The S-matrix for magnon-vortex scattering
has been considered (without the stabilizing Skyrme and potential
terms) in~\cite{ikw96} for antiferromagnets and
in~\cite{ismw98} for ferromagnets).
Because of the constraint $\Bphi \cdot \Bphi =1$ the model
effectively possesses two independent fields. 
We therefore introduce two--component small amplitude fluctuations
\be\label{fluct}
\mbox{\boldmath $\eta$} = \mbox{\boldmath $\hat r$}_n \eta_L
+ \mbox{\boldmath $\hat \varphi$}_n \eta_T \, , \qquad
\mbox{\boldmath $\hat r$}_n = \left( \begin{array}{c}
\cos n \varphi   \\ \sin n \varphi  \end{array} \right) \, , \quad 
\mbox{\boldmath $\hat \varphi$}_n = \left( \begin{array}{c}
-\sin n \varphi   \\ \cos n \varphi  \end{array} \right) 
\ee
which we decompose into components parallel and perpendicular
to the soliton configuration with winding number $n$. Of course,
there are many different ways to parametrize the total
time-dependent field $\Bphi$ which lead to different equations of
motion for the corresponding fluctuations. However, 
the bound--states and the scattering matrix, 
in particular the phase--shifts are independent of the chosen
parametrization. A very convenient choice
different from the common Polyakov parametrization~\cite{p75}, 
which only at first sight may seem complicated, is given by 
\be\label{ck}
\Bphi = \left( \begin{array}{c}
\mbox{\boldmath $\hat r$}_n \sin F (1-\mbox{\boldmath $\eta$}^2/2)   
+ \mbox{\boldmath $\hat r$}_n \cos F \eta_L
+ \mbox{\boldmath $\hat \varphi$}_n \eta_T  \\ 
\cos F (1-\mbox{\boldmath $\eta$}^2/2) - \sin F \eta_L
\end{array} \right) 
\stackrel{r \longrightarrow \infty}{\longrightarrow}
\left( \begin{array}{c} \mbox{\boldmath $\eta$} \\   
1-\mbox{\boldmath $\eta$}^2/2 \end{array} \right)  \, . 
\ee
The main advantage of this parametrization is that it leads to a flat
metric for the non--linear sigma model part of the lagrangian. 
For 3D-$SU(N)$ skyrmions this parametrization is well--known
as Callan--Klebanov ansatz~\cite{ck85}.

It is now straightforward to write down the e.o.m. for the
fluctuating components $\eta_L$ and $\eta_T$. The Lagrangian
(\ref{lag}) has to be expanded to quadratic order
in the fluctuations, then the e.o.m. can be read off. We give these
coupled linear differential equations explicitely, where we have
already exploited the time--dependence as well as the angular
dependence of the fluctuations 
\be\label{phonon}
\eta_L = \sum_M f_M (x) e^{iM\varphi}  e^{-i \omega t} \, , \qquad
\eta_T = i \sum_M g_M (x) e^{iM\varphi}  e^{-i \omega t}
\, .
\ee
The e.o.m. decouple in the magnetic quantum number $M$, which is the
analogon of the phonon- (or grand-) spin familiar from scattering
calculations for 3D-$SU(N)$ skyrmions
\bea\label{eom}
&&-\frac{1}{x} \left( x b_L f_M^\prime \right)^\prime 
+ \frac{M^2 }{x^2} f_M + \frac{n^2}{x^2} ( b_T-2s^2 ) f_M 
-a \frac{2n^2c}{x} \left( \frac{F^\prime s}{x} \right)^\prime f_M  
+ a c f_M \qquad 
\nonumber \\
&& \qquad \qquad -\frac{nMc}{x^2} ( 1+b_T ) g_M 
+ a \frac{nM F^\prime s}{x^2} g_M^\prime 
+ a \frac{2nM}{x} \left( \frac{F^\prime s}{x} \right)^\prime g_M    
= \omega^2 b_L f_M  \nonumber \\
&& -\frac{1}{x} \left( x g_M^\prime \right)^\prime 
+ \frac{M^2 }{x^2} b_T g_M 
- ( F^{\prime 2} - \frac{n^2 c^2}{x^2} ) g_M 
-a \frac{n^2c}{x} \left( \frac{F^\prime s}{x} \right)^\prime g_M 
+ a c g_M  \qquad \\
&& \qquad \qquad -\frac{nMc}{x^2} ( 1+b_T) f_M 
- a \frac{nM F^\prime s}{x^2} f_M^\prime 
+ a \frac{nM}{x} \left( \frac{F^\prime s}{x} \right)^\prime f_M
= \omega^2 b_T g_M \nonumber  
\, . 
\eea
Here we introduced the metric functions $b_L(x)=1+a n^2 s^2/x^2$
and $b_T(x)=1+a F^{\prime 2}$. The energies $\omega$ are
understood in natural units of $\sqrt{fem}$ such that the threshold occurs
at $\omega = \sqrt{a}$.
The equations are invariant with
respect to the simultaneous replacements $M \to -M$ and $g_M \to -g_M$.
These equations contain
all the information about zero--modes, bound--states, the scattering
matrix and the stability of the hedgehog solutions.

\subsection{Zero-modes}

The soliton's energy (\ref{static}) is invariant under spatial
rotations around the $z$-axis (which for the hedgehog 
is equivalent to an iso--rotation around the internal $3$-axis)
as well as under a translation in the
$x$-$y$ plane. The infinitesimal rotation (iso--rotation)
and translation give rise to zero--modes which are solutions
of the e.o.m. (\ref{eom}) for $\omega^2=0$.

The rotational zero--mode is obtained by a shift of the azimuthal
angle $\varphi \to \varphi + \alpha$ in the hedgehog ansatz
(\ref{hedgehog}). Comparison with (\ref{ck}) and (\ref{phonon})
shows then that this zero--mode 
\be\label{rot}
f_0(x)=0 \, , \quad g_0(x)=s \qquad \mbox{(rotational zero-mode)}
\ee
is located in the $M=0$ partial--wave.
Similarly, the translational zero--mode 
$\mbox{\boldmath $r$} \to \mbox{\boldmath $r$} + \mbox{\boldmath $a$}$
\be\label{trans}
f_1(x)=F^\prime \, , \quad g_1(x)=\frac{ns}{x} 
\qquad \mbox{(translational zero-mode)}
\ee
sits in the $M=1$ partial--wave.

With the stability condition (\ref{stability})
it is easily checked that the above modes satisfy the e.o.m. for
$\omega^2=0$. As both zero--modes possess a finite norm, they will act
as bound--states in the scattering calculation and influence the
phase--shifts via Levinson's theorem. Besides these
zero energy bound--states there are also bound--states at finite
energy.

\subsection{Bound-states}

In both systems, $n=1$ and $n=2$, we observe bound--states
in the $M=n$ partial--wave. With decreasing parameter $a$ the
energies of these bound--states move towards the threshold,
in particular for the $n=1$ system this happens very quickly
(cf. Table \ref{tab_2}).
There is a simple explanation of this phenomenon. As is noticed
from the e.o.m. the iso--rotation with respect to the $1$- and 
$2$-axes
\be\label{bound}
f_n(x)=1 \, , \quad g_n(x)=c 
\qquad (\mbox{zero-energy solution for} \quad a \to 0)
\ee
is a zero--energy solution of the $M=n$ partial--wave for $a \to 0$.
The potential connected with a finite $a \not= 0$ distorts this
continuum solution into a bound--state, which in the limit $a \to 0$
is shifted towards the threshold.
\begin{table}[htbp]
\begin{center}
\parbox{10.5 cm}{\caption{\label{tab_2}
Bound-state energies relative to the threshold 
in the $M=n$ partial--wave. With decreasing 
parameter $a$ the bound--states move towards threshold.
}} 
\begin{tabular}{|c|cccccc|}
\hline
$a$   & $1.0$   & $0.707$   & $0.316$       & $0.224$      & 
                               $0.1$         & $0.071$      \\        
\hline
$n=1$  & $0.943$ & $0.99994$ &  $\simeq 1.0$ & $\simeq 1.0$ &
                                $\simeq 1.0$ & $\simeq 1.0$ \\
$n=2$  & $0.330$ & $0.394$   &  $0.622$      & $0.690$      &
                                $0.866$      & $0.943$      \\
\hline
\end{tabular}
\end{center} 
\end{table}

In the $n=1$ system this is the only bound--state apart from the
zero--modes already discussed. However in the $n=2$ case some of the
low-lying resonances may also become bound for sufficiently strong $a$,
e.g. for $a=1$ we find additional bound--states in the $M=0$ and
$M=1$ partial--waves at energies $\omega_0=0.923$ and 
$\omega_1=0.793$ respectively. All these bound--states must be
considered for the Casimir energy (compare also the discussion
of the phase--shifts in subsection 3.4).

\subsection{Stability of hedgehog solitons}

Formally, eqs. (\ref{eom}) are the e.o.m. for small amplitude fluctuations
around a hedgehog soliton with arbitrary topological quantum number
$n$. For this reason they contain information about the stability of
the $n$-skyrmion. For the $n$-skyrmion to be stable there must not
exist solutions to (\ref{eom}) with negative $\omega^2$.

Indeed, we do find solutions of (\ref{eom}) with negative $\omega^2$
for topological quantum numbers $n \ge 3$. For example for $n=3$
with $a^2=0.1$ there exist such solutions in the 
$M=2$ ($\omega^2=-0.14$), $M=3$ ($\omega^2=-0.30$) and
all higher partial waves. In the $n=1$ and $n=2$ sectors 
there exist no negative $\omega^2$ solutions.

We conclude, that the $n=1$ and $n=2$ hedgehog solitons are stable
while those with higher topological charges are unstable. This
agrees with the findings in ref.~\cite{psz95}. In what follows we
consider the scattering off the stable skyrmions with $n=1$ and
$n=2$. 

\subsection{Phase--shifts}

For the $2$-dimensional scattering problem in spherical
coordinates~\cite{mf} the incident plane wave 
is decomposed into partial waves
carrying magnetic quantum numbers
\be\label{plane}
e^{i \mbox{\boldmath $pr$}} = \sum_{M=-\infty}^{\infty} i^M
J_M(pr) e^{iM \varphi} =
\sum_{M=0}^{\infty} \epsilon_M i^M J_M(pr) \cos (M \varphi) \, .
\ee
The latter transformation with the multiplicities $\epsilon_0=1$
and $\epsilon_M=2$ for $M \ge 1$
follows from the properties of the regular
and irregular Bessel functions of the first kind, $J_M(pr)$ and
$N_M(pr)$, when $M \to -M$. Thus, it suffices to consider the 
partial waves $M \ge 0$. Similarly, for the scattered wave
(plane wave plus outgoing radial wave) in the case of a single
channel we have
\bea\label{scatt}
\Psi &=& \sum_{M=0}^{\infty} \epsilon_M i^M \psi_M (p,r) 
\cos (M \varphi) 
\, .
\eea
The partial--wave projected scattering waves $\psi_M (p,r)$ with
appropriate boundary 
conditions at the origin are integrated according to the e.o.m. and
the phase--shifts then are obtained from the asymptotic form
\be\label{scatt1}
\psi_M (p,r) \to [ J_M(pr) \cos \delta_M(p) + N_M(pr) \sin \delta_M(p) ]
e^{i \delta_M(p)}
\, .
\ee
The phase--shifts are related to the cross--section in the usual way,
and the property $\delta_M(p)=\delta_{-M}(p)$ 
follows from the corresponding symmetry of the e.o.m.. The
generalization to two coupled channels now is straightforward.
It is noticed that our scattering equations (\ref{eom}) 
decouple asymptotically when the functions 
$f_{\pm}= (f_M \mp g_M) / \sqrt{2} $ are introduced
\be\label{asym}
-\frac{1}{r} \left(r f_\pm^\prime \right)^\prime 
+ \frac{(M \pm n)^2}{r^2} f_\pm = p^2 f_\pm \, , \qquad
\omega^2=p^2+m^2 \, ,
\ee
with the solution
\be\label{asym1}
f_{\pm} (p,r) = A_{\pm} (p) J_{|M \pm n|}(pr) 
+ B_{\pm} (p) N_{|M \pm n|}(pr)
\, .
\ee
From the coefficients $A_{\pm} (p)$ and $B_{\pm} (p)$ of the
asymptotical solution, the $2 \times 2$ scattering matrix $S_M$
is obtained for every partial wave $M$. Again, as the e.o.m. are
invariant with respect to $M \to -M$ so is the scattering matrix
and it suffices to consider $M \ge 0$. It is convenient to
diagonalize the scattering matrix
\be\label{smatrix}
S_M = U_M \left( \begin{array}{cc} 
e^{i\delta_M^{(1)}}     &    0   \\   
0        &    e^{i\delta_M^{(2)}}      \end{array} \right)
(U_M)^\dagger \, ,
\ee
in order to obtain the intrinsic eigen phase--shifts $\delta_M^{(1)}$
and $\delta_M^{(2)}$. Their sum $\delta_M = \delta_M^{(1)} +
\delta_M^{(2)}$ is plotted for the partial waves with $M \le 4$ 
in Fig. 3, where the $n=1$ and $n=2$ systems are considered separately.

The picture that emerges looks qualitatively quite similar to what
has been obtained for 3D-$SU(N)$ skyrmions long ago~\cite{we84}. 
For topological charge $n=1$ we obtain in detail:
The phases for $M=0$ and $M=1$, where the rotational and translational
zero modes are located together with the $M=1$ bound--state,
start at $\pi$ resp. $2\pi$ according to Levinson's theorem.
For the smaller parameters $a^2=0.01$ and $0.1$ the 
wave--function of the $M=1$
bound--state close to threshold (Table \ref{tab_2})
is already quite similar to the infinitesimal translation
(cf. subsection 3.5 for the limit $a \to 0$). 
As a consequence the phase--shift at threshold
reacts with a sudden drop from $2\pi$ to $\pi$ which in the
figures is only noticed because the phase--shift falls below
$\pi$ before it bends to the right. A weakly pronounced breathing
mode is observed at low energies for $M=0$. The 
partial--waves $M=2,3,\dots$ then contain a band of resonances
which are flattened out for higher $M$ and also with decreasing 
parameter $a$.

Similarly for topological charge $n=2$: For the smaller parameters
$a^2=0.01$ and $0.1$ the $M=0$ and $M=1$ phase--shifts start
at $\pi$ because of the zero modes. In all but the $M=2$ 
partial--wave with its bound--state (cf. Table \ref{tab_2})
sharp resonances occur, which are much more pronounced compared
to the $n=1$ case. For strong potentials then, the $M=0$ and $M=1$
resonances become bound (e.g. for $a=1$ at $\omega_0=0.923$
and $\omega_1=0.793$). Also less pronounced secondary resonances
do appear. 
\begin{figure}
\begin{center}
\begin{tabular}{cc}
\epsfig{figure=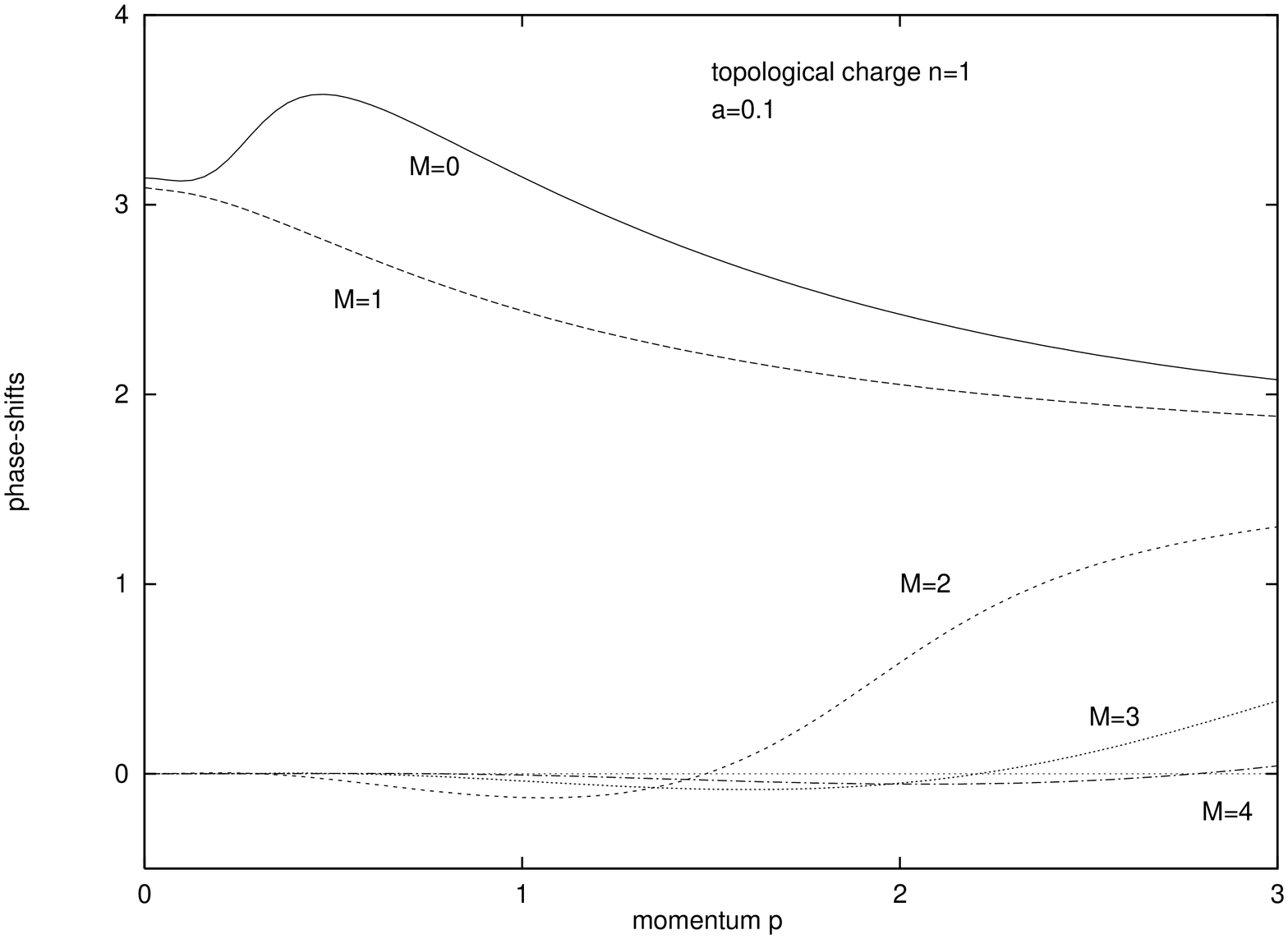,width=5cm,angle=270} &
\epsfig{figure=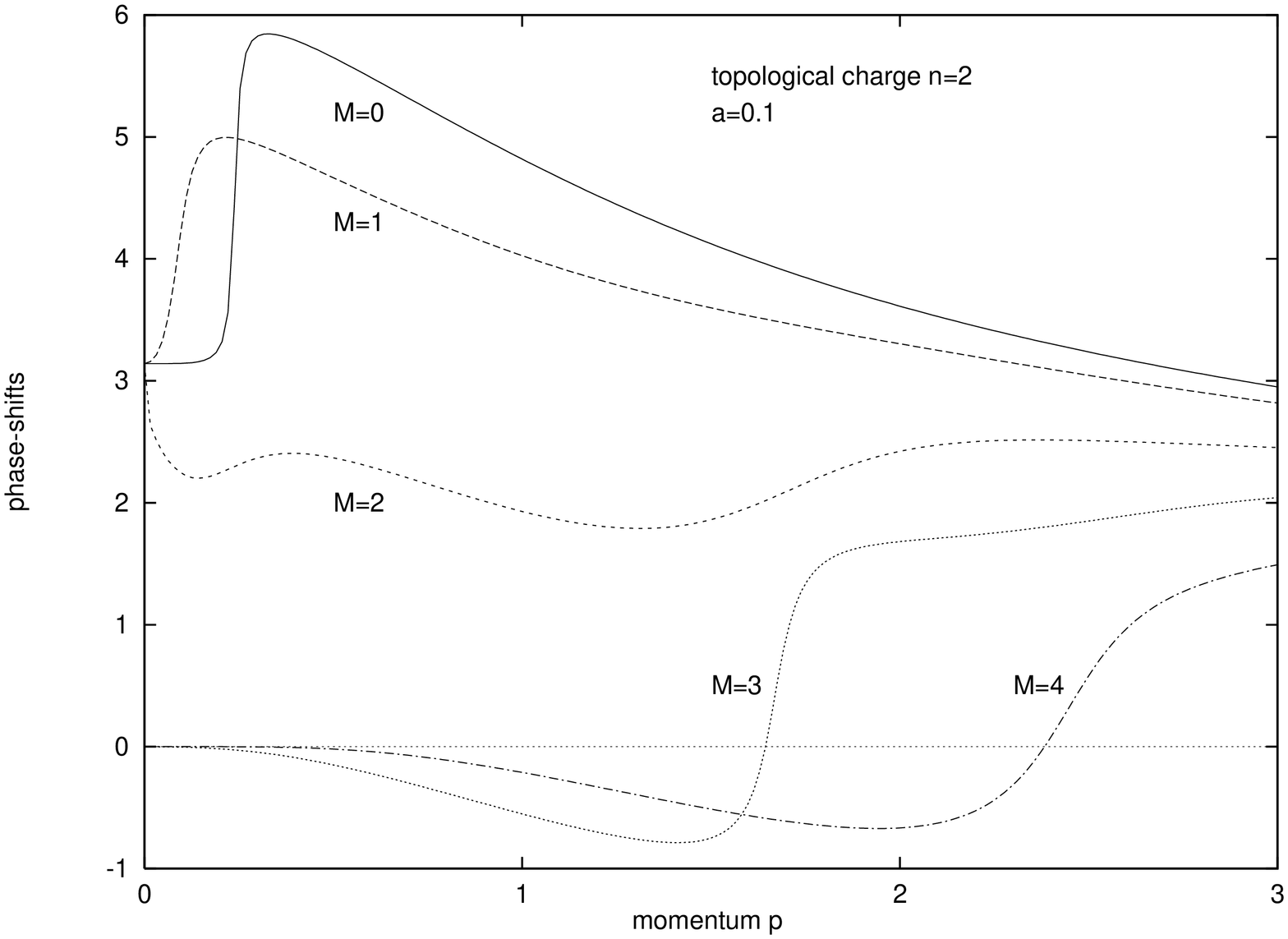,width=5cm,angle=270} \\
\epsfig{figure=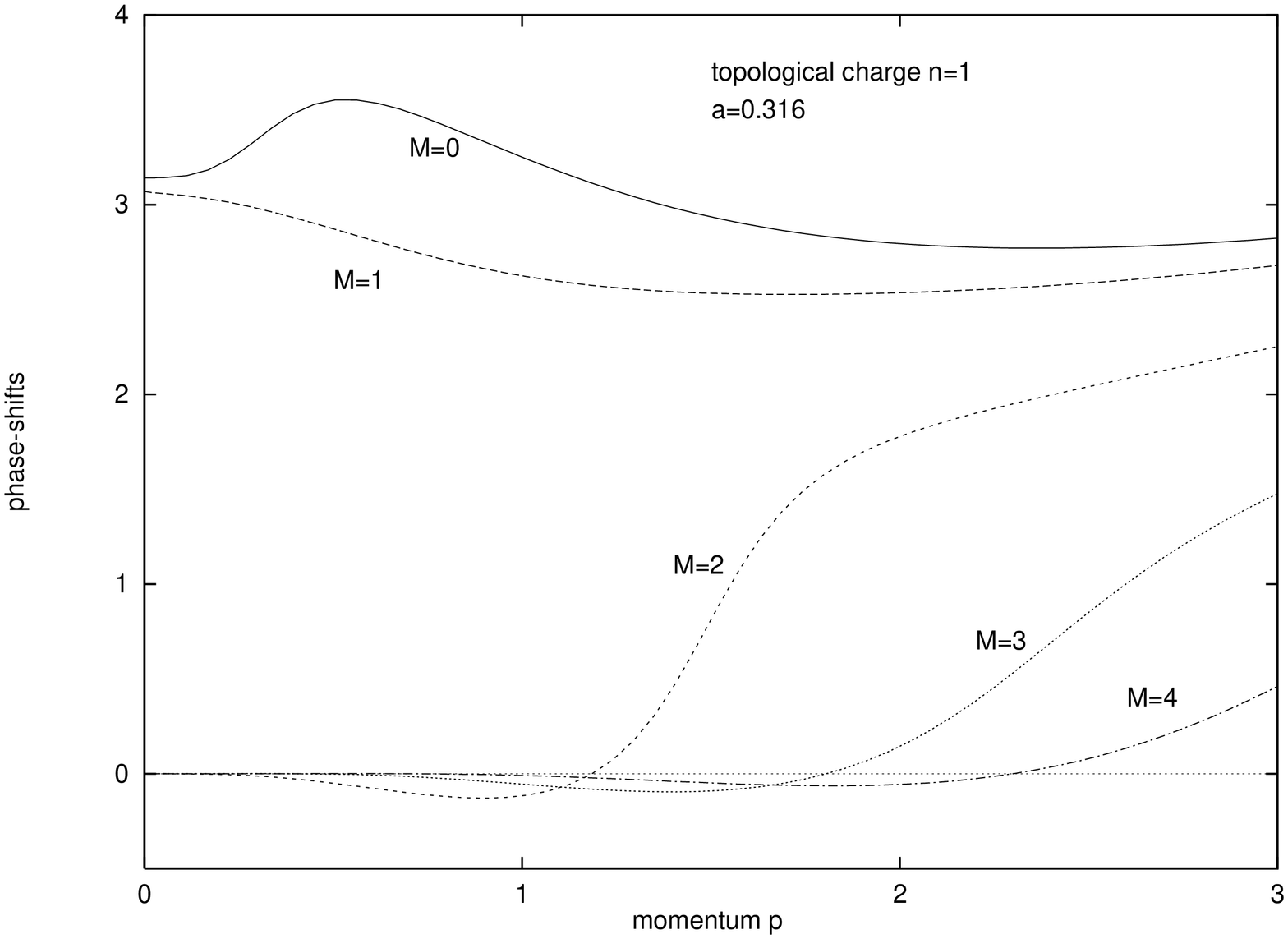,width=5cm,angle=270} &
\epsfig{figure=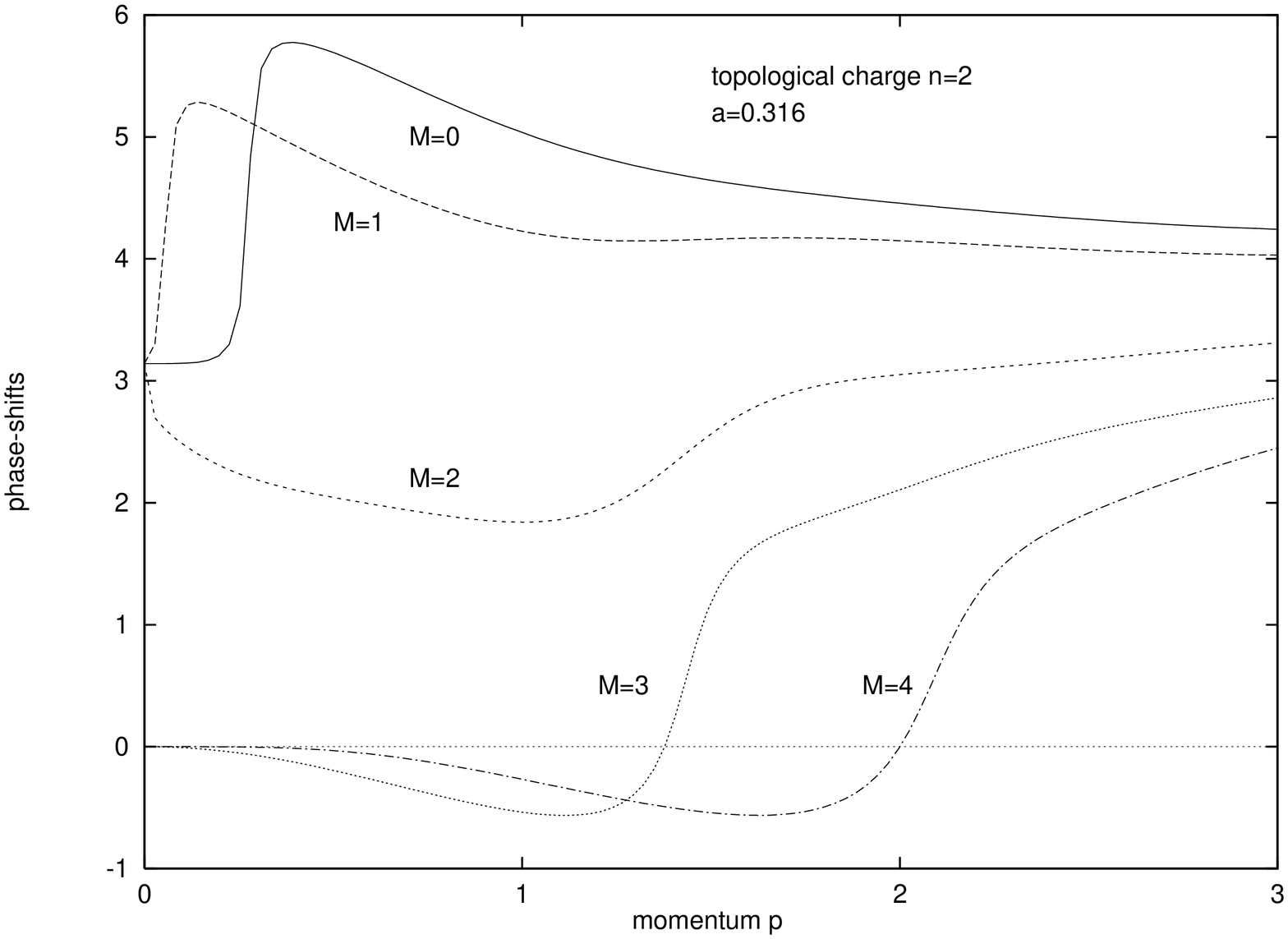,width=5cm,angle=270} \\
\epsfig{figure=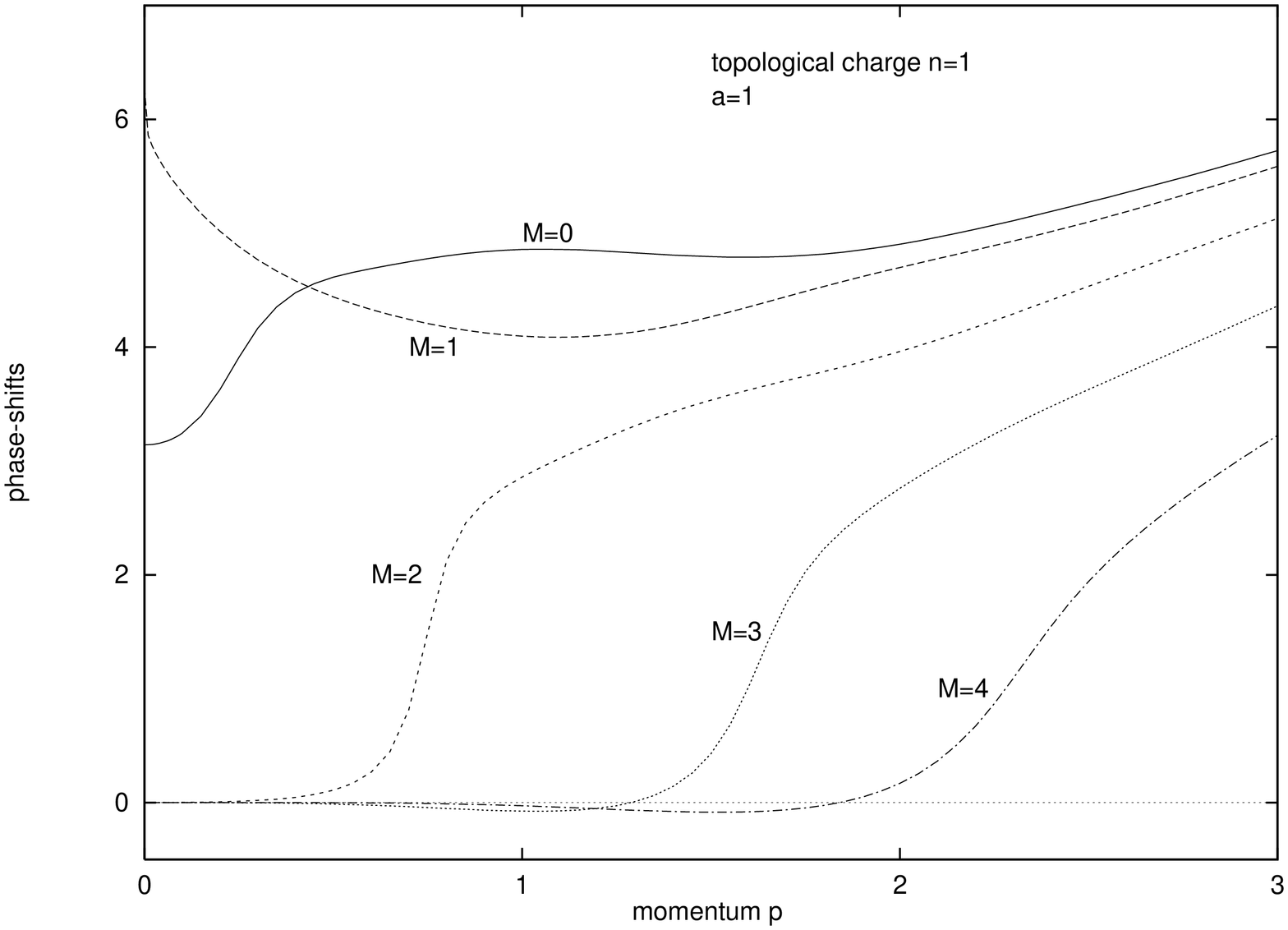,width=5cm,angle=270} &
\epsfig{figure=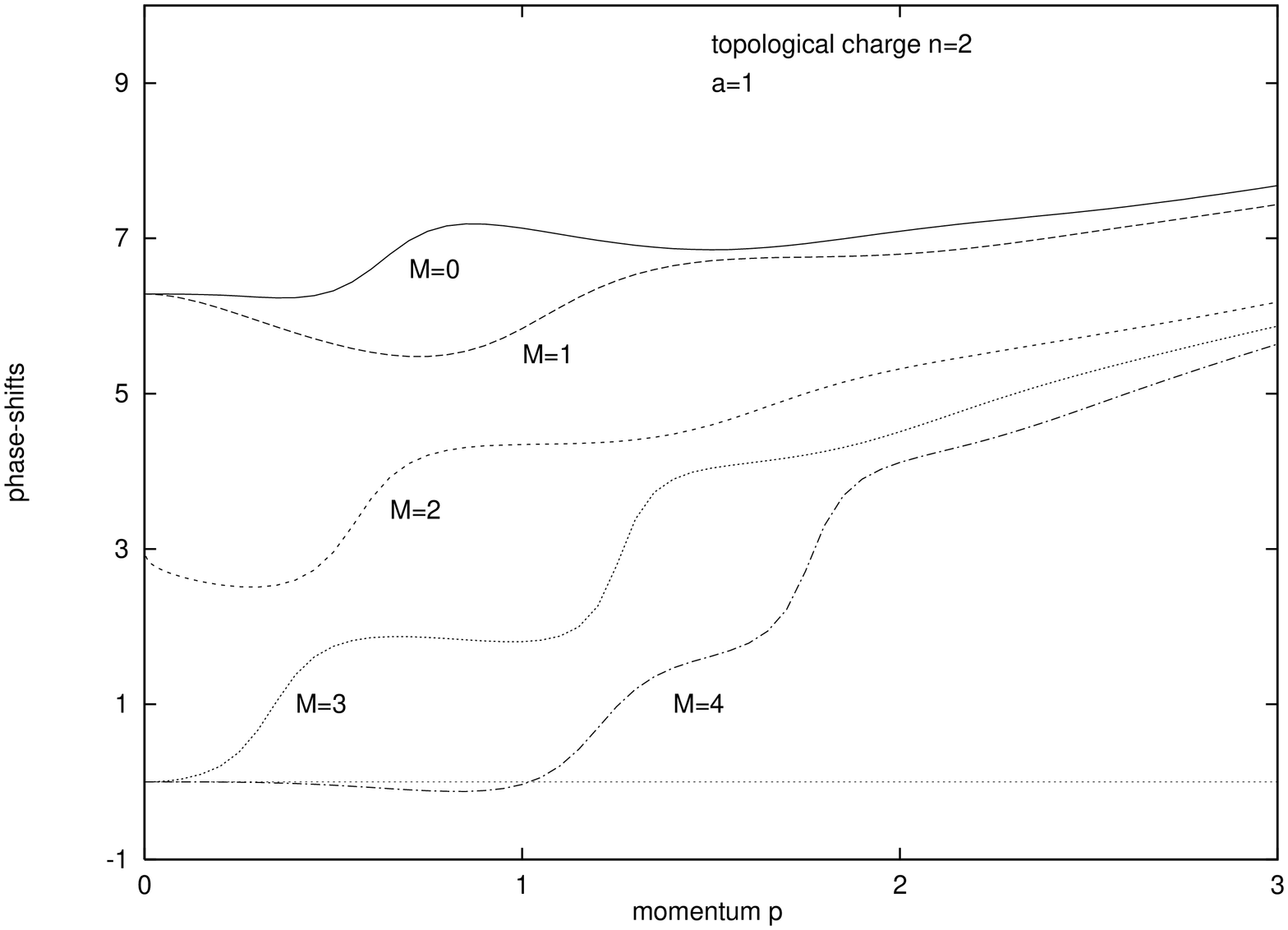,width=5cm,angle=270} 
\end{tabular}
\protect\caption{Phase shifts for the lowest partial waves
scattered off the stable $n=1$ and $n=2$ skyrmions.
The relevant parameter was fixed at $a^2=0.01$, $0.1$ and $1.0$.
The momentum $p$ is plotted in natural units.
According to Levinson's theorem the phase--shifts start at
multiples of $\pi$. Many resonances are observed in the various
channels.
}
\end{center}
\end{figure}

\subsection{Belavin-Polyakov soliton}

Finally we would like to add a few comments on the pure 
Belavin--Polyakov solution~\cite{bp75}. 
As mentioned we recover this solution
for small parameters $a$ with its size $R$ still fixed by the
balance of the Skyrme and potential terms (cf. appendix). In the
pure N$\ell \sigma$ model without additional stabilizing terms
the size $R$ of the BP soliton is an undetermined parameter.
In contrast to the full lagrangian (\ref{lag}) the pure
N$\ell \sigma$ model does not break $O(3)$--symmetry and is also
conformally invariant. Due to these additional symmetries a larger
number of zero--modes is expected.

The e.o.m. for the fluctuations obtained in that case from (\ref{eom})
with $a=0$ and $x=r/R$ decouple for $f_{\pm}= (f_M \mp g_M) / \sqrt{2}$
(not only asymptotically) and are easily solved. Still the counting of the
zero--modes is a bit tricky. In addition to the rotation
(\ref{rot}), the breathing--mode (conformal invariance) becomes a
zero--mode in the $M=0$ partial--wave as expected. But for the
1--soliton the iso--rotation (\ref{bound}) coincides with the
translation (\ref{trans}) giving rise to only $2 \times 1$
zero--modes in the $M=1$ partial--wave. This makes altogether
$4$ zero--modes for the 1--soliton. Similar counting for the
2--soliton yields $8$ zero--modes.

Of course, this discussion is somewhat academic, because in reality
there should always be some stabilizing term which fixes the size
of the soliton and which at least breaks conformal invariance.
Therefore we do not show the corresponding phase--shifts.

\section{Casimir energy}

With the bound--state energies and the phase--shifts provided in
the previous section we are in the position to evaluate the
1-loop contribution to the soliton energy, i.e. the Casimir energy.
The UV singularities contained in the loop are related to the
high momentum behaviour of the phase--shifts
\be\label{psum}
\delta (p) = \sum_{M=0}^\infty \epsilon_M \delta_M (p) 
\stackrel{p \longrightarrow \infty}{\longrightarrow}
a_0 p^2 + a_1
\, .
\ee
In the case of a vanishing Skyrme term the coefficients
\be\label{coeff}
a_0=0 \, , \qquad a_1=\frac{1}{4} \int d^2 r \left[ F^{\prime 2} 
+ \frac{n^2 s^2}{r^2} + 2 m^2 (1-c) \right]
\ee
are known analytically from the Born terms. 
For the full model they have to
be extracted numerically from the phase--shift sum (\ref{psum}),
which for that purpose has to be calculated to a high precision
with some $100$ partial waves taken into account. 
Typical values obtained, e.g. for $a^2=0.1$, are $a_0=1.0 (fem)^{-1}$
and $a_1=7.2$ for the 1-soliton, and $a_0=2.0 (fem)^{-1}$
and $a_1=13.4$ for the 2-soliton respectively.
According to (\ref{coeff})
the N$\ell \sigma$ plus potential term contributions to these
coefficients are $a_1=8.3$ for $n=1$ and $a_1=15.6$ for $n=2$.
With this established we may use the phase--shift formula~\cite{dhn74},
subtract the troublesome high momentum behaviour from the phase--shifts,
and separately add the corresponding counterterms
\bea\label{1-loop}
E^{1-loop} \! &=& \! \frac{1}{2\pi} 
\left[ -\int_0^\infty d p \frac{p}{\omega}
\delta (p) - m \delta (0) \right] + \frac{1}{2} \sum_B \omega_B
\\
\! &=& \! -\frac{1}{2\pi} \int_0^\infty d p \frac{p}{\omega}
\left( \delta (p) - a_0 p^2 - a_1 \right) 
+ \frac{1}{2} \sum_B (\omega_B - m)  
- \int \frac{d^2p}{(2\pi)^2} \, \frac{a_0p^2+a_1}{\omega}  
\, . \nonumber
\eea
This
procedure is closely related to what has been employed in the $3+1$
dimensional Skyrme model for the calculation of the Casimir energy to the
nucleon mass~\cite{m93,mw97}. 
The counter terms could e.g. be evaluated using a 2-momentum cutoff
$\Lambda >> m$
\be\label{cut}
\int \frac{d^2p}{(2\pi)^2} \, \frac{1}{\omega} =
\frac{1}{2\pi} [\Lambda -m] \, , \qquad 
\int \frac{d^2p}{(2\pi)^2} \, \frac{p^2}{\omega} =
\frac{1}{6\pi} [\Lambda^3 - \frac{3}{2} m^2 \Lambda + 2m^3] \, ,
\ee
which makes the linear and cubic divergencies explicit
\be\label{loop}
E^{1-loop} =
-\frac{a_0}{6\pi} \left[ \Lambda^3 - \frac{3}{2} m^2 \Lambda \right] 
- \frac{a_1}{2\pi} \Lambda + E^{cas}
\, .
\ee
While the N$\ell \sigma$ and the potential term contributions
located in $a_1$ may be absorbed in a renormalized constant
\be\label{fren}
f^2 = \bar f^2 - \frac{\Lambda}{4\pi}
\ee
($\bar f$ denotes the bare constant), the divergencies stemming
from the Skyrme term renormalize the Skyrme parameter $e$ as well
as the couplings of all the higher gradient terms not listed in 
the lagrangian (\ref{lag}). The latter renormalization we do not
carry out explicitely, instead we simply assume that the renormalized
Skyrme parameter be $e$ and the renormalized couplings of the
higher gradient terms be zero. To which extent this is a consistent
assumption will be discussed in the following section in connection
with the scale--dependence of our results. Now all the divergencies
residing in (\ref{loop}) are taken care of, leaving the finite
Casimir energy
\be\label{casimir}
E^{cas}= -\frac{1}{2\pi} \int_0^\infty d p \frac{p}{\omega}
\left( \delta (p) - a_0 p^2 - a_1 \right) 
+ \frac{1}{2} \sum_B (\omega_B - m)  
- \frac{a_0 m^3}{3\pi} + \frac{a_1 m}{2\pi} 
\, .\ee
This result is independent of the employed regularization scheme.
For instance, using dimensional regularization the counter
terms (\ref{cut}) in $2+1$ dimensions become finite and equal 
to the last terms in the brackets, respectively, 
which yields the same expression for the Casimir energy. 
However, although dimensional regularization in odd
dimensions has been used right from the beginning~\cite{bg72} and
is now commonly applied in $\Phi^4_3$, QCD$_3$ and 
Chern-Simons theories~\cite{bg86,m90}, 
we hesitate to apply this scheme for an odd number of dimensions
because the fate of the UV singularities remains obscure.
As the divergencies do not show up in odd dimensional
regularization there seems to be no possibility to introduce
a renormalization scale. For strictly (super) renormalizable theories 
this may be of no further importance, however in theories which are
renormalized order by order in terms of a gradient expansion a scale
is desirable in order to control the convergence of the series 
(cf. section 5).

\begin{figure}[htbp]
\centerline{\epsfig{figure=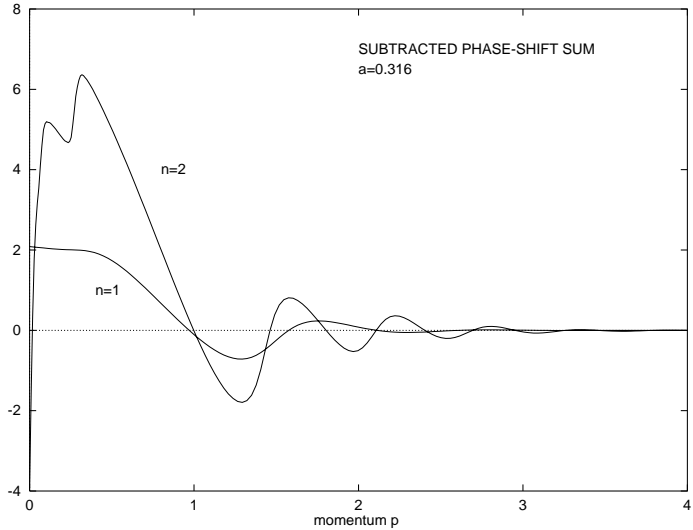,width=7cm,angle=270}}
\protect\caption{Subtracted phase--shift sum
as a function of the momentum $p$ in natural units
for the $n=1$ and $n=2$ systems calculated with $a^2=0.1$.
}
\end{figure}
The Casimir energy (\ref{casimir}) consists 
of three parts, i.e. the phase--shift integral, the
bound--state contribution and the contribution from the 
counter--terms.

The subtracted phase--shift sum $\delta (p) - a_0 p^2 - a_1$ which
enters the phase--shift integral is plotted in Fig. 4 for the
$n=1$ and $n=2$ systems ($a^2=0.1$). The maximum momentum to
which this sum has to be integrated depends sensitively on the
parameter $a$. 
The various contributions to the Casimir energy are given in
Table \ref{tab_3} for several values of $a$.
\begin{table}[htbp]
\begin{center}
\parbox{12cm}{\caption{\label{tab_3}
Individual contributions to the Casimir energy according
to (\ref{casimir}) for the $n=1$ and $n=2$ systems. All
energies are given in natural units. 
}}
\begin{tabular}{|c|rrr|rrr|}
\hline
   &&  $n=1$ &&& $n=2$ & \\
\hline
$a$  & $0.1$   &  $0.316$  &  $1.0$ & $0.1$   &  $0.316$  &  $1.0$ \\ 
\hline
phase-shifts           
   & $-0.29$  & $-0.08$ & $0.00$ & $-0.63$  & $-0.26$ & $0.05$  \\
bound-states  
   & $-0.47$ & $-0.84$ & $-1.56$ & $0.52$ & $-1.06$ & $-2.47$  \\
counter terms
   & $0.31$  & $0.62$  & $1.85$  & $0.61$  & $1.16$  & $3.36$  \\
\hline
total
   & $-0.45$  & $-0.30$ & $0.29$ & $-0.54$  & $-0.15$ & $1.00$  \\
\hline
\end{tabular}
\end{center} 
\end{table}
The dependence of the Casimir energies 
on this parameter is shown in Fig. 5. Note that for an easier
comparison again only half of the 2-skyrmion's Casimir energy is
plotted. 
\begin{figure}[htbp]
\centerline{\epsfig{figure=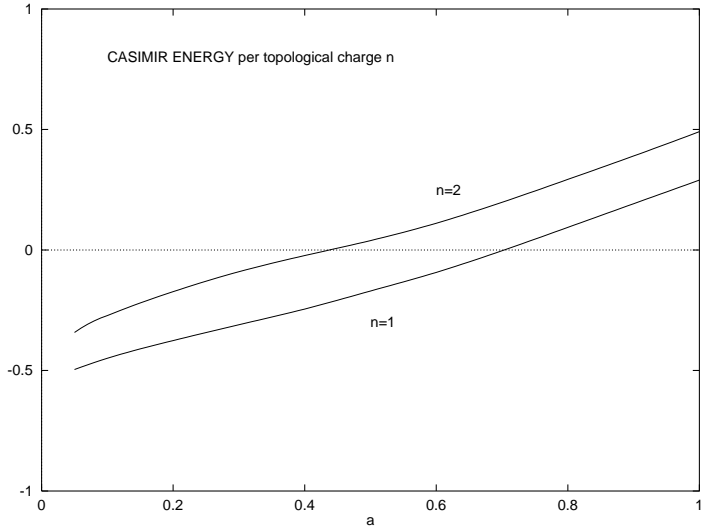,width=7cm,angle=270}}
\protect\caption{Casimir energies in natural units
for the 1- and 2-skyrmion
devided by their topological charge.
}
\end{figure}
It is noticed that the Casimir energy of the 1-skyrmion stays always
smaller compared to half of that of the 2-skyrmion. Therefore, in
contrast to the classical energy the Casimir energy favors single
skyrmions. These competing effects may be summarized in the formula
for the total energy cast into the form 
\bea\label{total}
E_n^{total}&=&4\pi f^2 E_n^0(a) +  \sqrt{fem}  E_n^1(a)
\nonumber \\  
&=& 4\pi f^2 \left[ E_n^0(a) +   y \,  E_n^1(a) \right]
\, , \qquad \qquad  y=\frac{\sqrt{fem}}{4\pi f^2}
\, , 
\eea
where $E_n^1(a)$ represents the Casimir energy (\ref{casimir})
in natural units $\sqrt{fem}$.
Thus, apart from the overall scale $4\pi f^2$ the model is
characterized by the two dimensionless parameters $a$ and $y$.
If this ratio $y$ exceeds a critical value (which 
increases slowly with $a$,
e.g. $y=0.3, 0.4, 0.7$ for $a^2=0.01, 0.1, 1.0$), then
$E_2^{total}-2E_1^{total}$ becomes positive and the 2-skyrmion
decays into two single 1--skyrmions.

\subsection{Numerical example from hadron physics}

Although different dimensionalities may cause qualitative
differences, let us illustrate these results by a numerical example
using the scale of hadron physics:
Take the value $a=0.1$, fix the size scale at
$1/\sqrt{fem}=0.2 fm$  ($\sqrt{fem} = 1$ GeV) and make use
of the numbers given in Table 1 and Table 3.
With the above parameters the 1-soliton gets a
topological radius $0.27$ fm and a Casimir energy
$E_1^{cas}=-0.45$GeV, a situation very similar to what has been
obtained in the hadronic Skyrme model~\cite{mw97}, if in addition
the classical soliton mass is fixed at $E_1^{class}=1.21 (4\pi f^2)
=1.39$ GeV in order to obtain the nucleon mass at
$E_1^{total}=0.94$ GeV. Then the $n=2$ soliton's total energy
$E_2^{total}=(2.64 - 0.53)$ GeV $= 2.11$ GeV 
turns out to be larger than that of two separated 1-skyrmions. 
The parameter in this example, 
$y = 0.9$, lies far above the critical value $0.3$.
Of course, the $2+1$ dimensional model discussed
here cannot be taken literally for hadron
physics, but exactly this mechanism may shift the
undesired torus configuration obtained in hadronic soliton models
with baryon number $B=2$ to higher energies.

\subsection{Casimir energy of the pure Belavin-Polyakov soliton}

Finally we would like to discuss the Casimir energy of the pure
BP soliton with free size parameter $R$ in order
to make contact to the quantum corrections as obtained in
ref.~\cite{r89}. With $m=0$ and the coefficients $a_0=0 \, , 
\, a_1=2\pi n$ and no additional bound--states present (cf. the
discussion about the zero--modes in subsection 3.5) eq.(\ref{1-loop})
for the 1--loop contribution simplifies
\bea\label{bp}
E^{1-loop} \! &\stackrel{BP}{=}& \! 
- \frac{1}{2\pi} \int_0^{(\Lambda)} d p \delta (p) 
\nonumber \\
\! &=& \! -\frac{1}{2\pi} \int_0^{(\Lambda)} d p 
\left( \delta (p) - a_1 \right)   - \frac{a_1}{2\pi} \Lambda  
\\
\! &=& \! -\frac{1}{2\pi R} \int_0^{(\Lambda R)} d (pR) 
\left( \delta (pR) - 2 \pi n \right)   - n \Lambda  
\, . \nonumber
\eea
We indicate here the possibility to limit the momentum integration
by a finite Debye momentum $\Lambda=p_D$ related to the lattice constant.
Because the soliton should at least cover several lattice
sites, $\Lambda R > 1$ is large enough to extend
the integration over the subtracted phase--shifts 
to infinity (see Fig. 4) just as in our field theoretical treatment.
Through the substitution in the last step the dependence of the subtracted
phase--shift integral on the soliton size $R$ becomes explicit:
the magnitude of the Casimir energy decreases with increasing
soliton size.

Numerically, we find
\be 
E^{cas}_1=-\frac{0.5}{R_1} \, , \qquad \qquad 
E^{cas}_2=-\frac{1.7}{R_2} 
\ee
for the pure BP solitons with topological charges $n=1$ and $n=2$
respectively. Their sizes $R_1$ and $R_2$ are arbitrary and generally
different. If the sizes are
determined by the stabilizing terms in the limit $a \to 0$,
we obtain $R_n=b_n/\sqrt{fem}$ with $b_1 \to 0$ and $b_2 \to \sqrt{2}$
(appendix). For that reason 
the Casimir energy of the 1-soliton plotted in Fig. 5 bends to negative
values and finally diverges as $a$ approaches zero. The Casimir energy
of the 2-soliton stays finite, 
$E^1_2 = -1.7/\sqrt{2} + {\cal O} ( \sqrt{a} )$, but with an infinite
slope at $a=0$. 

In ref.~\cite{r89} the subtracted phase--shift integral was neglected for
$p_D R >> 1$ and the last
term in (\ref{bp}) then lead to the result 
$E^{1-loop} \stackrel{BP}{=} -n p_D$. Because this is exactly the
term we absorb into a renormalized $f^2$ (\ref{loop},\ref{fren}), 
this result corresponds
to $E^{cas}=0$ in our notation. The opposite limit $p_D R << 1$
where the 1-loop contribution (\ref{bp}) vanishes
corresponds to soliton sizes much smaller than the lattice spacing
and does not make sense. The conclusion, that the quantum corrections
lower the soliton energy as soliton size increases, is wrong.

\section{Scale dependence}

In this section we introduce a scale $\mu$ which allows to shift finite
pieces from the tree to the 1-loop contribution and vice versa.
It will be introduced in such a way
that we recover the results of the previous section for 
$\mu = \mu_0$. Tentatively, $\mu_0=4\pi f^2$ may be identified with
the underlying energy scale of the model in 
analogy to the chiral scale $\mu_0=4\pi f_\pi \simeq 1$GeV of
ChPT~\cite{e96}. Our results will however be presented in such a way
that no fixation of $\mu_0$ is required.
For this purpose we write the cutoff characterizing the divergencies
of the 1-loop contribution (\ref{loop}) as a sum of two parts
$\Lambda = [\Lambda - (\mu-\mu_0)] +(\mu-\mu_0)$. The first part
is then renormalized into a scale-dependent strength of the
N$\ell \sigma$ and potential term
\be\label{fren1}
f^2 (\mu) = \bar f^2 - \frac{\Lambda - (\mu-\mu_0)}{4\pi}
= f^2 + \frac{\mu-\mu_0}{4\pi} \, . 
\ee
According to this formula $\mu \not= \mu_0$ shifts
the energy scale $4 \pi f^2(\mu)$ away from its original value
$4 \pi f^2$. In order to exhibit the scale--dependence it is
now convenient to present our results as function of
the ratio $f^2(\mu)/f^2$ rather than $\mu$ itself.
The presentation
is then independent of $\mu_0$ and also robust
against changes in the regu\-larisation scheme 
which may introduce numerical factors to the scale (e.g. in the
3-momentum instead of the 2-momentum cutoff scheme 
the scale is changed by a factor of $\pi/2$).
The second part of the above decomposition, $(\mu-\mu_0)$,
remains explicit in the finite Casimir energy

\bea\label{casimir1}
E^{cas} (\mu) &=& -\frac{1}{2\pi} \int_0^\infty d p \frac{p}{\omega}
\left( \delta (p) - a_0 p^2 - a_1 \right) 
+ \frac{1}{2} \sum_B (\omega_B - m) \\  
&& + \frac{a_0}{6\pi} \left[ (\mu-\mu_0)^3 -
\frac{3m^2}{2}(\mu-\mu_0)-2m^3 \right]
- \frac{a_1}{2\pi} \left[ \mu - \mu_0 - m \right] 
\, . \nonumber
\eea
Also for the $a_0$ - term with its cubic divergence 
we kept all contributions from the second part 
in the Casimir energy although other prescriptions are possible.
However this ambiguity will not influence the results strongly
because of the smallness of $a_0$ (cf. numbers in the previous
section). In this way both, the tree  and the 1-loop contributions
become scale--dependent. While for the N$\ell \sigma$ and potential
terms the scale--dependence in the total soliton energy is exactly
compensated, the 1-loop contribution which arises from
the Skyrme term would not only yield a scale--dependent Skyrme
parameter $e(\mu)$, but would also switch on all higher gradient
terms assumed to be zero for $\mu=\mu_0$. Having assumed all
these couplings to be independent of the scale implies
that the tree + 1-loop contribution to the soliton energy
cannot be strictly scale--invariant. In fact, the resulting 
scale--dependence measures the magnitude of the higher gradient terms not
accounted for. All the more it comes as a surprise, just as in the
hadronic case~\cite{mw97}, that for specific parameter choices an
almost scale--independent soliton mass may be obtained. Such a case
($a^2=0.1, y=1$) is depicted in Fig.6 for the 1-soliton. The rather
strong scale--dependence of the tree contribution is nicely
compensated over a wide range $f^2 (\mu)=0.5 f^2, \dots , 2.0 f^2$
when the 1-loop contribution is added.
Of course, this statement has to depend on the parameters used.
Therefore in Fig. 7 we plotted the total 1-soliton mass for $a^2=0.1$
and various values of the ratio $y$. It is noticed that weak
scale--dependence requires a ratio close to $y \simeq 1$,
the value used
in the previous figure. 
\begin{figure}[htbp]
\centerline{\epsfig{figure=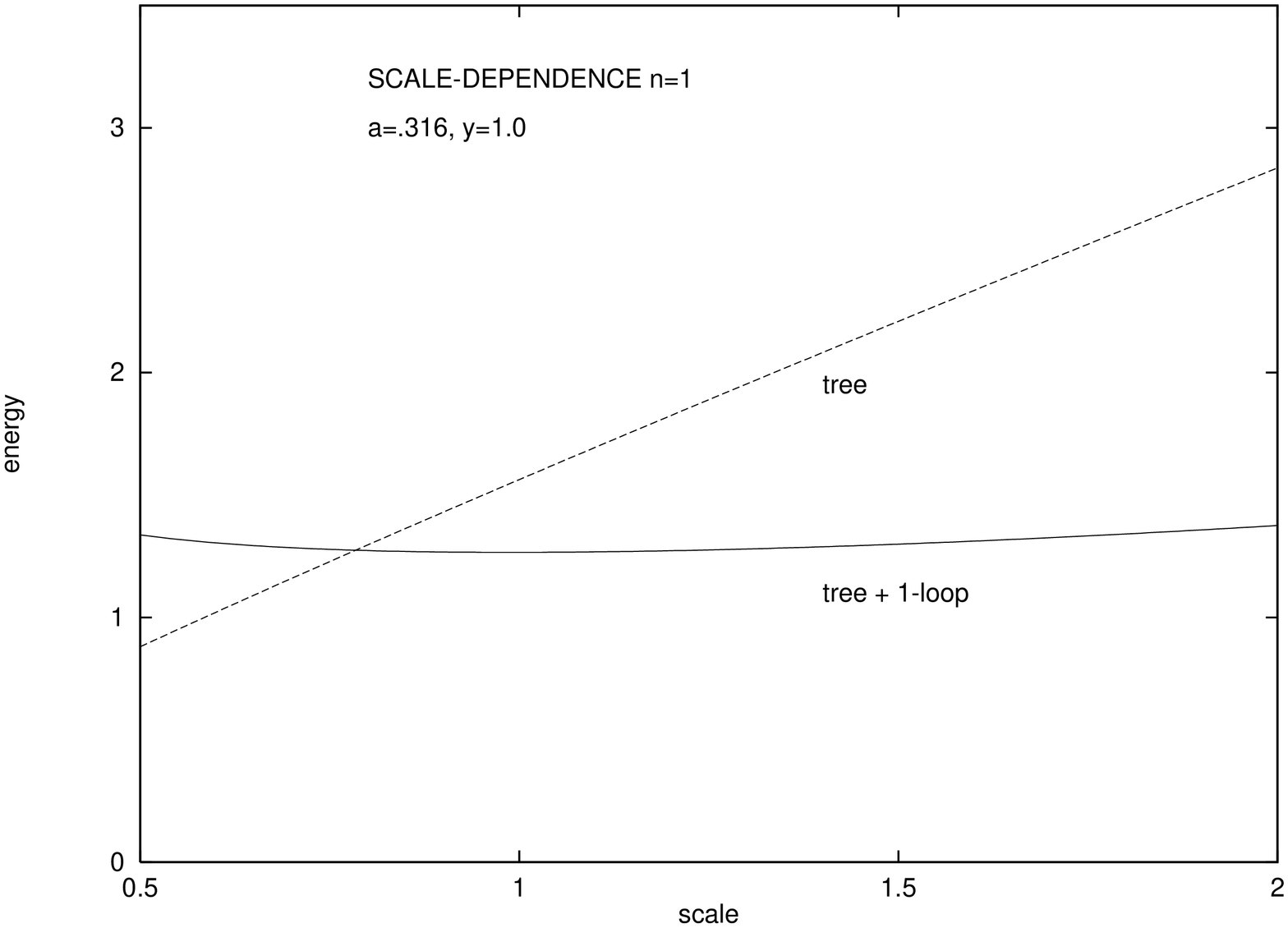,width=7cm,angle=270}}
\protect\caption{Scale-dependence of the 1-soliton's energy
in tree and tree + 1-loop. The dimensionless parameters are
$a=a(\mu_0)=.316$ and $y=y(\mu_0)=1.0$. The energy is 
in units $4\pi f^2$ and the scale is given by the ratio
$f^2(\mu)/f^2$.
}
\centerline{\epsfig{figure=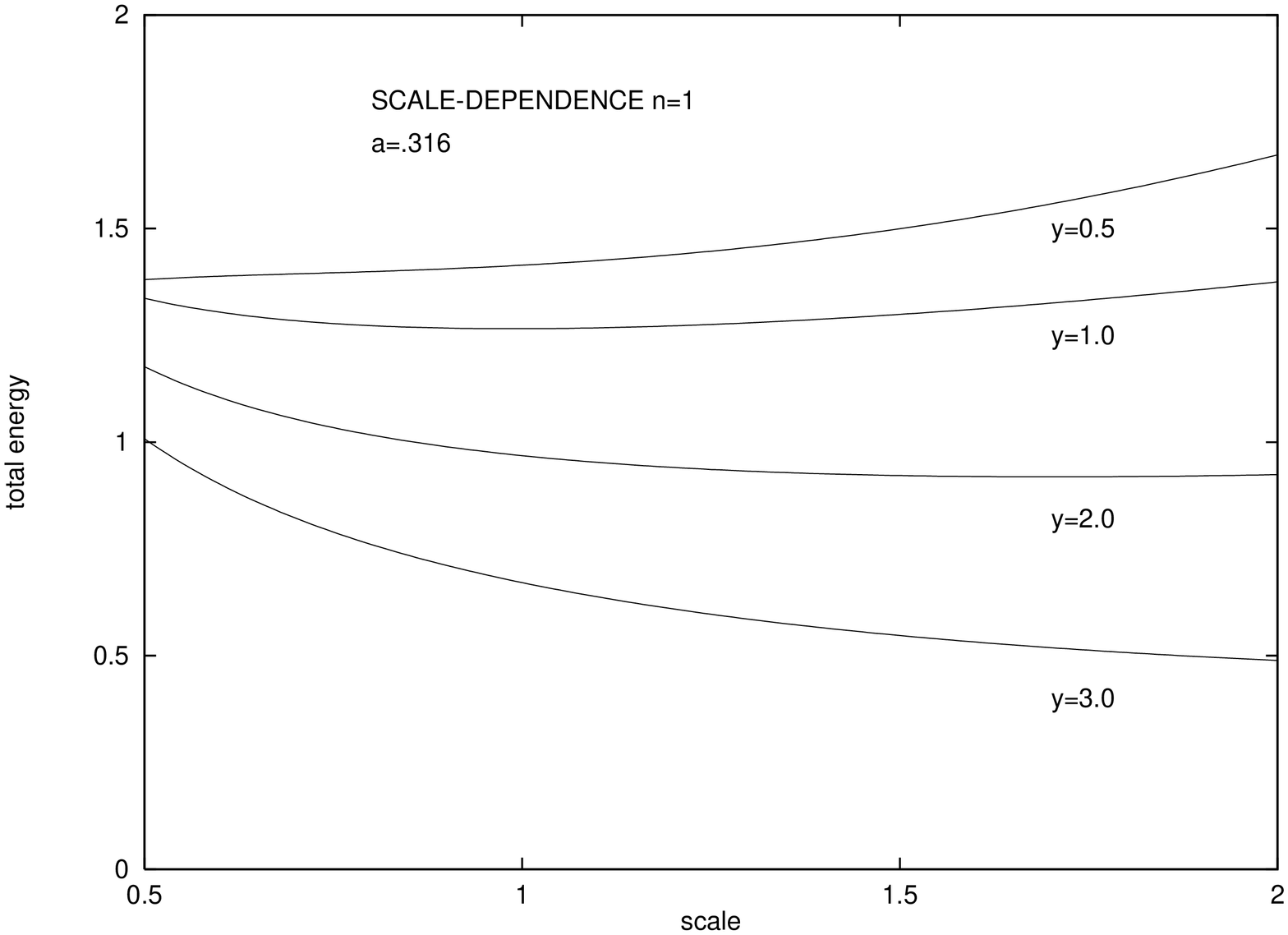,width=7cm,angle=270}}
\protect\caption{Scale-dependence of the 1-soliton's energy
in tree + 1-loop for $a=a(\mu_0)=.316$ and 
various parameters $y=y(\mu_0)$. 
Energy and scale as in Fig. 6.
}
\end{figure}
Lower and higher values of $y$ enhance the
scale--dependence. We do not show the corresponding plots for the
2-soliton because they look quite similar, with the scale--dependence
being even somewhat weaker in that case.

We may now pose the question for which parameter combinations $a$ and
$y$ we may expect a modest scale--dependence in accordance with the
assumption that the couplings of all higher gradient terms are zero.
The answer is illustrated in Fig.8. 
\begin{figure}[htbp]
\centerline{\epsfig{figure=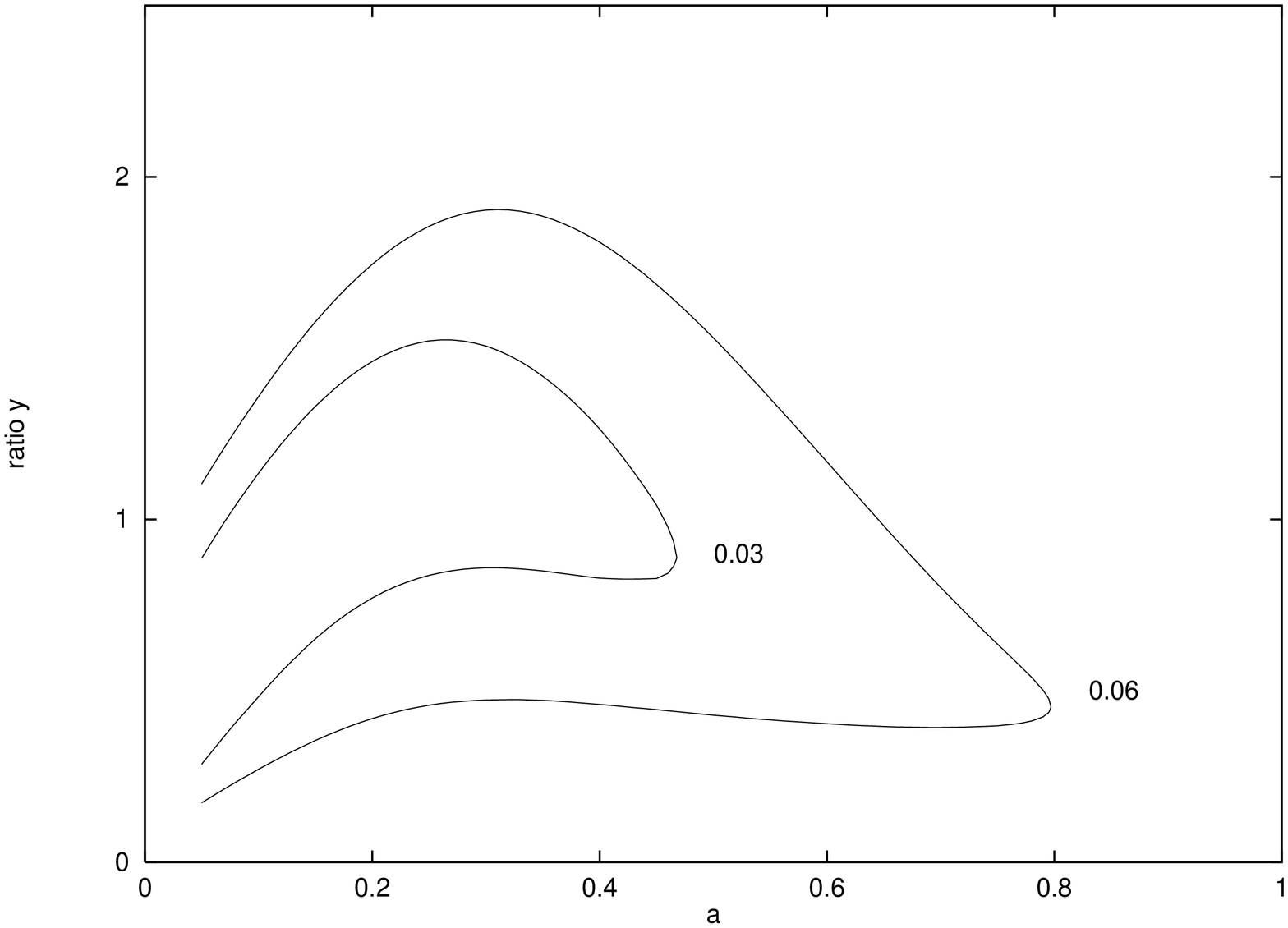,width=7cm,angle=270}}
\protect\caption{Parameter space $(a,y)$ for 1-skyrmions.
Inside the inner contour the scale--dependence is less than $3\%$,
inside the outer contour less than $6\%$.
}
\end{figure}
Inside the inner contour the average scale--dependence in the intervall
$f^2(\mu) / f^2 \in [0.5,2]$ is less than $3\%$. For comparison
the $6\%$ contour is also plotted. 

All the parameter combinations lying inside the contours 
yield negative Casimir energies whose absolute values though sizeable
do not exceed $50\%$ of the classical soliton mass. They also
lead to an
unstable 2-soliton which decays into two individual 1-solitons.
In particular the parameters $(a=0.1, y=0.9)$ of our numerical example
from hadron physics (section 4.1) lie in the
center of the almost scale--independent region.

The weak scale dependence obtained
for parameter combinations lying inside the contours in Fig. 8
implies that loop--corrections 
can be reliably calculated within the framework
of the lagrangian (\ref{lag}) with all higher couplings set 
equal to zero.

\section{Conclusions}

We studied the magnon--vortex system in the $2+1$ dimensional
$O(3)$ model in a field theoretical approach. For that purpose the
N$\ell \sigma$ model was augmented by stabilizing standard fourth--order
Skyrme and potential terms.

Complete information on bound and scattering states in all
partial--waves was established. We find an extremely rich excitation
spectrum with pronounced resonances as in the 3D-$SU(N_f)$ case with
its baryon resonances. In principle these resonances should be
accessible by measuring the excitation spectrum of spin--waves
in the presence of skyrmions.

Furthermore, the quantum fluctuations will allow to study their
influence on the shape 
of the soliton~\cite{a97}.
Note that in tree approximation the
third component of the order parameter field is always
tied to $-1$ at the origin for topological reasons,
in variance with microscopic Hartree-Fock calculations
for the ferromagnet~\cite{HF97}
which suggest a vanishing value in case of small solitons. 

Finally with the bound state energies and the scattering
phase--shifts the Casimir energies were evaluated. The
appearing UV singularities were renormalized under the
assumption that the (renormalized) couplings of the higher
gradient terms be small. A criterium for the consistence of this 
assumption is the approximate scale--independence
of the results such that the effective action represents
an almost 1-loop renormalizable theory. This criterium requires a detailed
balance of tree and 1-loop contributions 
which limits the parameter space where this requirement is met.
The following results apply for this
restricted parameter space. Parameters lying outside
that range may lead to other conclusions, but there the 
1-loop contribution cannot be reliably calculated within the model.
\begin{itemize}
\item  The Casimir energy is negative and generally large leading
       to a considerable reduction of the total soliton energy.
       However, it does not exceed $50\%$ of
       the tree contribution and thus may still be considered 
       a (sizeable) correction.
\item  The magnitude of the Casimir energy decreases with increasing
       soliton size. But it also decreases as the stabilizing terms
       become more important relative to the N$\ell \sigma$ model
       term.
\item  With the 1-loop corrections included the $n=2$ soliton becomes
       unstable and decays into two individual $n=1$ solitons (which
       may still be weakly bound by a dipole force). We conjecture
       that the same situation might occur for the 3D-$SU(2)$ 
       $B=2$ torus in hadron physics.
\end{itemize}
It is obvious that the program outlined in this paper has to be
tailored for specific applications concerning ferromagnets and
antiferromagnets. Most importantly for ferromagnets the
time derivative part of the lagrangian has to be replaced by the
$T$ violating Landau-Lifshitz term with one time derivative only.
This replacement may change the results for the Casimir energies
appreciably. Also, for antiferromagnets an external magnetic field 
should couple through the time component of the
covariant derivative. Finally, proper consideration of the
non--local Coulomb interaction complicates the situation: instead
of coupled differential equations coupled integro--differential
equations have to be solved for the fluctuations.

The evaluation of loop corrections as presented here naturally suggest 
the inclusion of temperature into the formalism. This opens the
possibility to study the properties of 2D spin--textures at finite 
temperature, which is of particular interest near the
symmetry restoring phase--transition.

\bigskip
\noindent
This work is supported in parts by funds provided by the FCT, Portugal
(Contract PRAXIS/4/4.1/BCC/2753).

\section*{Appendix}

In this appendix we present analytical soliton solutions for 
the two limiting cases $a \to 0$ and $a \to \infty$.

\subsubsection*{Belavin-Polyakov solution for $a \to 0$}

The analytical solution of the Euler-Lagrange equation
(\ref{stability}) in the limit $a \to 0$
$$
\frac{1}{x} \left( x F^\prime \right)^\prime - \frac{n^2 s c}{x^2} =0
\eqno{({\rm A}.1)}
$$
with boundary conditions (\ref{boundary}) is the well--known
Belavin-Polyakov soliton
$$
\tan \frac{F}{2} = \left( \frac{b_n}{x} \right)^n \, .
\eqno{({\rm A}.2)}
$$
The size parameter $b_n^2 = \sqrt{ 2n(n^2-1)/3 }$
is determined by the interplay of the Skyrme- and potential terms
contributing to the static energy functional
$$
E_n^0 = n + a \, \frac{\pi}{3b_n^2} \, \frac{n^2-1}{\sin \pi/n }
      + a \, \frac{b_n^2}{2} \, \frac{\pi/n}{\sin \pi/n }
\, , \qquad n \not= 1 \, .
\eqno{({\rm A}.3)}
$$
The corresponding square radius follows from its definition
(\ref{radius})
$$
< x^2 >_n = \sqrt{\frac{2n}{3} (n^2-1) } \, \frac{\pi/n}{\sin \pi/n }
\, \qquad n \not= 1 \, .
\eqno{({\rm A}.4)}
$$
It is noticed that in case of the 1-soliton the contribution
from the potential term to ({\rm A}.3) diverges. For that reason
we modify ({\rm A}.2),
$$
\tan \frac{F}{2} = b_n \,
\sqrt{ \frac{\epsilon}{e^{\epsilon x^2}-1}} \, ,
\eqno{({\rm A}.5)}
$$
such that for $\epsilon \to 0$ the Belavin--Polyakov solution is
recovered. The variation of the static energy functional in lowest
order $\epsilon$
$$
E_1^0 = 1 + \frac{\epsilon}{2} + \frac{2a}{3b_1^2}
- \frac{ab_1^2}{2} \ell n \epsilon 
\eqno{({\rm A}.6)}
$$
gives then $\epsilon = ab_1^2$ and $\ell n \epsilon =  - 4 / 3b_1^4$ 
with the result
$$
E_1^0 = 1 + \epsilon ( \frac{1}{2} - \ell n \epsilon ) \, , \qquad \qquad
a^2 = - \frac{3}{4} \epsilon^2 \ell n \epsilon \, .
\eqno{({\rm A}.8)}
$$
Indeed with $a$ also $\epsilon$ tends to zero. The square radius
behaves like
$$
<x^2>_1 = \sqrt{-\frac{4}{3} \ell n \epsilon }
\eqno{({\rm A}.9)}
$$
and diverges weakly in the limit $a \to 0$. For
$a \stackrel{<}{_\sim} 0.001$ the radius of the 1-soliton exeeds
that of the 2-soliton. Exactly this behaviour may be traced in the
numerical solution if the differential equation (\ref{stability})
is solved for very small $a$ with great precision. Because the
square radius ({\rm A}.9) equals the derivative of the soliton
energy with respect to the parameter $a$, this also explains the
infinite slope at the origin indicated in the 1-soliton's curve
plotted in Fig. 2.

\subsubsection*{Solution for $a \to \infty$}

The Euler-Lagrange equation (\ref{stability}) in the opposite
limit $a \to \infty$
$$
\frac{n^2 s}{x} \left( \frac{F^\prime s}{x} \right)^\prime - s = 0
\eqno{({\rm A}.10)}
$$
may also be solved analytically
$$
\cos F(x) = \frac{1}{8n^2} (b_n^2 - x^2) x^2 + 2 \frac{x^2}{b_n^2} - 1
\, , \qquad x \le b_n \, .
\eqno{({\rm A}.11)}
$$
The integration constants are fixed by the condition $F(b_n)=0$
connected with topological charge $n$. The size parameter $b_n^2=4n$
again minimizes the static energy functional
$$
E_1^0 = na \left[ \frac{b_n^2}{4n} + 
\frac{1}{2} \left( \frac{4n}{b_n^2} \right)
- \frac{1}{6} \left( \frac{b_n^2}{4n} \right)^3 \right] 
\eqno{({\rm A}.12)}
$$
with the results quoted in (\ref{limit})
$$
E_n^0 = \frac{4n}{3} a \, , \qquad \qquad
<x^2>_n =  \frac{4n}{3} \, .
\eqno{({\rm A}.13)}
$$
Needless to say, that numerical solutions of the
Euler-Lagrange equation (\ref{stability}) reproduce these
results for large enough $a$.

\end{document}